# On the relativistic two-point Green's function

A. D. Alhaidari

*Physics Department, King Fahd University of Petroleum & Minerals, Box 5047,
Dhahran 31261, Saudi Arabia*
E-mail: **haidari@mailaps.org**

Using a recently developed approach for solving the three dimensional Dirac equation with spherical symmetry, we obtain simple representations for the Green's function of the Dirac-Oscillator and Dirac-Coulomb problems. This is accomplished by setting up the relativistic problem in such a way that makes comparison with the nonrelativistic problem highly transparent and results in a map of the latter into the former. An advantage of this representation is in the computational economy as depicted by the number of terms, which are quadratic in the Whittaker functions. These are single terms except for the off-diagonal elements of the Dirac-Coulomb Green's function which, nonetheless, simplifies into the sum of a first order and third order relativistic terms.

**Introduction**: Recently, an effective approach has been developed for solving the Dirac equation for spherically symmetric local interactions. It was applied successfully to the solution of various relativistic problems [1-6]. These include, but not limited to, the Dirac-Coulomb, Dirac-Morse, Dirac-Scarf, Dirac-Pöschl-Teller, Dirac-Woods-Saxon, etc. The central idea in the approach is to separate the variables such that the two coupled first order differential equations resulting from the radial Dirac equation generate Schrödinger-like equations for the two spinor components. This makes the solution of the relativistic problem easily attainable by simple and direct correspondence with well-known exactly solvable nonrelativistic problems. There are two main ingredients in the formulation of the approach that make it work. The first is a global unitary transformation of the Dirac equation which, of course, reduces to the identity in the nonrelativistic limit. The second is the introduction, in a natural way, of an extra potential component which is constrained to depend, in a particular way, on the independent potential function of the problem.

The main objective in all previous applications of the approach was in obtaining the discrete energy spectrum and its associated spinor eigenfunctions [1-3,6]. In this article, however, we demonstrate how one can utilize the same approach in generating the two-point Green's function – an important object of prime significance in the calculation of relativistic physical processes. One of the main findings here is in obtaining the relativistic Green's function for the Dirac-Oscillator which, to the best of our knowledge, has not been calculated before. Another contribution is in the Dirac-Coulomb problem where the representation of the diagonal elements of the Green's function has the advantage of being constructed as single terms quadratic in the Whittaker functions. However, the off-diagonal elements are not, but they are reduced into two additive terms: a first order, as well as, third order relativistic terms. The implications, if any, on the efficiency of theoretical calculations due this simplification have yet to be made and assessed.

We start by setting up the relativistic problem following the procedure in the above-mentioned approach and then give the solution for the particle propagator in the Dirac-Oscillator and Dirac-Coulomb potentials. In atomic units ($m = \hbar = 1$) and taking the speed of light $c = \lambdabar^{-1}$, we write the Hamiltonian for a Dirac spinor coupled to a four-component potential $(A_0, \vec{A})$ as follows:

$$H = \begin{pmatrix} 1 + \lambdabar A_0 & -i\lambdabar \vec{\sigma}\cdot\vec{\nabla} + i\lambdabar \vec{\sigma}\cdot\vec{A} \\ -i\lambdabar \vec{\sigma}\cdot\vec{\nabla} - i\lambdabar \vec{\sigma}\cdot\vec{A} & -1 + \lambdabar A_0 \end{pmatrix} \quad (1)$$

where $\lambdabar$ is the Compton wavelength scale parameter $\hbar/mc$ and $\vec{\sigma}$ are the three 2×2 Pauli matrices. It is to be noted that this type of coupling does not support an interpretation of $(A_0, \vec{A})$ as the electromagnetic potential unless, of course, $\vec{A} = 0$ (e.g., the Coulomb potential). That is, the wave equation with this Hamiltonian is not invariant under the usual electromagnetic gauge transformation. Imposing spherical symmetry and writing $(A_0, \vec{A}) = [\lambdabar V(r), \hat{r} W(r)]$, where $\hat{r}$ is the radial unit vector, gives the following two component radial Dirac equation

$$\begin{pmatrix} 1 + \lambdabar^2 V(r) - \varepsilon & \lambdabar\left[\frac{\kappa}{r} + W(r) - \frac{d}{dr}\right] \\ \lambdabar\left[\frac{\kappa}{r} + W(r) + \frac{d}{dr}\right] & -1 + \lambdabar^2 V(r) - \varepsilon \end{pmatrix} \begin{pmatrix} f^+(r) \\ f^-(r) \end{pmatrix} = 0 \quad (2)$$

where $\varepsilon$ is the relativistic energy and $\kappa$ is the spin-orbit quantum number defined as $\kappa = \pm (j + \frac{1}{2})$ for $\ell = j \pm \frac{1}{2}$. $V(r)$ and $W(r)$ are real radial functions referred to as the even and odd components of the relativistic potential, respectively. Eq. (2) results in two coupled first order differential equations for the two radial spinor components. Eliminating one component in favor of the other gives a second order differential equation. This will not be Schrödinger-like (i.e., it contains first order derivatives) unless $V = 0$. To obtain Schrödinger-like equation in the general case we proceed as follows. A global unitary transformation $\mathcal{U}(\eta) = \exp(\frac{i}{2}\lambdabar\eta\sigma_2)$ is applied to the Dirac equation (2), where $\eta$ is a real constant parameter and $\sigma_2$ is the 2×2 matrix $\begin{pmatrix} 0 & -i \\ i & 0 \end{pmatrix}$. The Schrödinger-like requirement relates the two potential components by the linear constraint $V = \pm(S/\lambdabar)[W + \kappa/r]$, where $S = \pm \sin(\lambdabar \eta)$. This results in a Hamiltonian that will be written in terms of only one arbitrary potential function; either the even potential component $V(r)$ or the odd one $W(r)$. The unitary transformation together with the constraint map Eq. (2) into the following one, which we choose to write in terms of the odd potential component [5]

$$\begin{pmatrix} C - \varepsilon + (1\pm 1)\lambdabar S\left(W + \frac{\kappa}{r}\right) & \lambdabar\left[-\frac{S}{\lambdabar} + C\left(W + \frac{\kappa}{r}\right) - \frac{d}{dr}\right] \\ \lambdabar\left[-\frac{S}{\lambdabar} + C\left(W + \frac{\kappa}{r}\right) + \frac{d}{dr}\right] & -C - \varepsilon - (1\mp 1)\lambdabar S\left(W + \frac{\kappa}{r}\right) \end{pmatrix} \begin{pmatrix} \phi^+(r) \\ \phi^-(r) \end{pmatrix} = 0 \quad (3)$$

where $C = \cos(\lambdabar \eta)$ and $\begin{pmatrix} \phi^+ \\ \phi^- \end{pmatrix} = \mathcal{U}\begin{pmatrix} f^+ \\ f^- \end{pmatrix}$. This gives the following equation for one spinor component in terms of the other

$$\phi^{\mp}(r) = \frac{\lambdabar}{C \pm \varepsilon}\left[\mp\frac{S}{\lambdabar} \pm CU + \frac{d}{dr}\right]\phi^{\pm}(r) \quad (4)$$

where $U(r) = W(r) + \kappa/r$. On the other hand, the resulting Schrödinger-like wave equation for the two spinor components reads

$$\left[-\frac{d^2}{dr^2} + C^2 U^2 \mp C\frac{dU}{dr} \pm \frac{2S\varepsilon}{\lambdabar}U - \frac{\varepsilon^2 - 1}{\lambdabar^2}\right]\phi^{\pm}(r) = 0 \quad (5)$$

In all relativistic problems that have been successfully tackled so far, Eq. (5) is solved by correspondence with well-known exactly solvable nonrelativistic problems [1-2,6]. This correspondence results in two parameter maps (one for each spinor component) relating the relativistic to the nonrelativistic problem. Using these maps and the known solutions (energy spectrum and wavefunctions) of the nonrelativistic problem one can easily and directly obtain the relativistic energy spectrum and spinor wavefunctions.

Now to the issue at hand – the Green's function. The relativistic 4×4 two-point Green's function $G(\vec{r},\vec{r}',\varepsilon)$ satisfies the inhomogeneous matrix wave equation $(H-\varepsilon)G = -\lambdabar^2 \delta(\vec{r}-\vec{r}')$, where the energy $\varepsilon$ does not belong to the spectrum of $H$. For problems with spherical symmetry, the 2×2 radial component $\mathcal{G}_\kappa(r,r',\varepsilon)$ of $G$ satisfies $(H_\kappa - \varepsilon)\mathcal{G}_\kappa = -\lambdabar^2 \delta(r-r')$, where $H_\kappa$ is the radial Hamiltonian operator in Eq. (3). It should be noted that our definition of the radial component of the Green's function differs by a factor of $(rr')^{-1}$ from other typical definitions. We write $\mathcal{G}_\kappa$ as

$$\mathcal{G}_\kappa(r,r',\varepsilon) = \begin{pmatrix} \mathcal{G}_\kappa^{++} & \mathcal{G}_\kappa^{+-} \\ \mathcal{G}_\kappa^{-+} & \mathcal{G}_\kappa^{--} \end{pmatrix} \tag{6}$$

where $\mathcal{G}_\kappa(r,r',\varepsilon)^\dagger = \mathcal{G}_\kappa(r',r,\varepsilon)$. Let $\Phi = \begin{pmatrix} \phi^+ \\ \phi^- \end{pmatrix}$ and $\overline{\Phi} = \begin{pmatrix} \overline{\phi}^+ \\ \overline{\phi}^- \end{pmatrix}$ be the regular and irregular solutions (at the origin) of Eq. (5), respectively. Using these two solutions, $\mathcal{G}_\kappa$ could be constructed as

$$\mathcal{G}_\kappa(r,r',\varepsilon) = \frac{1}{\Omega_\kappa(\varepsilon)} \left[ \theta(r'-r)\Phi(r,\varepsilon)\overline{\Phi}^\top(r',\varepsilon) + \theta(r-r')\overline{\Phi}(r,\varepsilon)\Phi^\top(r',\varepsilon) \right] \tag{7}$$

where $\theta(r'-r)$ is the Heaviside unit step function and $\Omega_\kappa(\varepsilon)$ is the Wronskian of the regular and irregular solutions:

$$\Omega_\kappa(\varepsilon) = \lambdabar^{-1}\Phi^\top(r,\varepsilon)\begin{pmatrix} 0 & 1 \\ -1 & 0 \end{pmatrix}\overline{\Phi}(r,\varepsilon) = \lambdabar^{-1}\left[\phi^+(r,\varepsilon)\overline{\phi}^-(r,\varepsilon) - \phi^-(r,\varepsilon)\overline{\phi}^+(r,\varepsilon)\right] \tag{8}$$

which is independent of $r$ as can be verified by differentiating with respect to $r$ and using Eq. (4). Eq. (7) results in the following expressions for the elements of $\mathcal{G}_\kappa$:

$$\mathcal{G}_\kappa^{\pm\pm}(r,r',\varepsilon) = \frac{1}{\Omega_\kappa(\varepsilon)} \phi^\pm(r_<,\varepsilon)\overline{\phi}^\pm(r_>,\varepsilon) \tag{9}$$

$$\mathcal{G}_\kappa^{\pm\mp}(r,r',\varepsilon) = \frac{1}{\Omega_\kappa(\varepsilon)} \left[ \theta(r'-r)\phi^\pm(r,\varepsilon)\overline{\phi}^\mp(r',\varepsilon) + \theta(r-r')\phi^\mp(r',\varepsilon)\overline{\phi}^\pm(r,\varepsilon) \right] \tag{10}$$

where $r_< (r_>)$ is the smaller (larger) of $r$ and $r'$. The equations satisfied by these elements are obtained from $(H_\kappa - \varepsilon)\mathcal{G}_\kappa = -\lambdabar^2 \delta(r-r')$. They parallel Eqs. (4) and (5) for $\phi^\pm$ and read as follows:

$$\left[ -\frac{d^2}{dr^2} + C^2 U^2 \mp C\frac{dU}{dr} \pm \frac{2S\varepsilon}{\lambdabar}U - \frac{\varepsilon^2-1}{\lambdabar^2} \right]\mathcal{G}_\kappa^{\pm\pm}(r,r',\varepsilon) = -(C\pm\varepsilon)\delta(r-r') \tag{11}$$

$$\mathcal{G}_\kappa^{\mp\pm}(r,r',\varepsilon) = \frac{\lambdabar}{C\pm\varepsilon}\left[\mp\frac{S}{\lambdabar} \pm CU + \frac{d}{dr}\right]\mathcal{G}_\kappa^{\pm\pm}(r,r',\varepsilon) \tag{12}$$

Using the exchange symmetry $r \leftrightarrow r'$ of $\mathcal{G}_\kappa$, Eq. (12) could be rewritten as a linear combination of the $\mathcal{G}_\kappa^{\pm\pm}$ terms with two real coefficients adding up to unity. That is, we write

$$\mathcal{G}_\kappa^{-+}(r,r',\varepsilon) = \mathcal{G}_\kappa^{+-}(r',r,\varepsilon) = \xi \frac{\hbar}{C+\varepsilon}\left[-\frac{S}{\hbar} + CU(r) + \frac{d}{dr}\right]\mathcal{G}_\kappa^{++}(r,r',\varepsilon)$$
$$+ (1-\xi)\frac{\hbar}{C-\varepsilon}\left[+\frac{S}{\hbar} - CU(r') + \frac{d}{dr'}\right]\mathcal{G}_\kappa^{--}(r,r',\varepsilon) \qquad (13)$$

where $\xi$ is an arbitrary real dimensionless parameter. This development will now be applied to our two problems.

**Dirac-Oscillator Green's function**: The objective of adding a potential, which is linear in the coordinate, to the Dirac equation in an analogy to the kinetic energy term which is linear in the momentum lead Moshinsky and Szczepaniak to the solution of the Dirac-Oscillator problem [7]. The nonrelativistic limit reproduces the usual Harmonic oscillator. The linear potential had to be added to the odd part of the Dirac operator resulting in a potential coupling which is a special case of that given in the Hamiltonian of Eq. (1) above. Subsequently, the Dirac-Oscillator attracted a lot of attention in the literature [8]. Our contribution here is to find its two-point Green's function using the tools of the approach given above. In this setting, the Dirac-Oscillator is the system described by Eq. (3) with $\eta = 0$ (i.e., $S = 0$, $C = 1$) and $W(r) = \omega^2 r$, where $\omega$ is the oscillator frequency. Therefore, Eq. (11) for the diagonal elements of the radial Green's function reads:

$$\left[-\frac{d^2}{dr^2} + \frac{\kappa(\kappa\pm 1)}{r^2} + \omega^4 r^2 + \omega^2(2\kappa\mp 1) - \frac{\varepsilon^2-1}{\hbar^2}\right]\mathcal{G}_\kappa^{\pm\pm}(r,r',\varepsilon) = -(1\pm\varepsilon)\delta(r-r') \qquad (14)$$

We compare this equation with that for the nonrelativistic radial Green's function $g_\ell(r,r',E)$ of the three dimensional isotropic oscillator:

$$\left[-\frac{d^2}{dr^2} + \frac{\ell(\ell+1)}{r^2} + \omega^4 r^2 - 2E\right]g_\ell(r,r',E) = -2\delta(r-r') \qquad (15)$$

where $\ell$ is the angular momentum quantum number and $E$ is the nonrelativistic energy. The comparison gives the following two maps between the relativistic and non-relativistic problems. The map concerning $\mathcal{G}_\kappa^{++}$ is

$$\begin{aligned} g_\ell &\to 2\mathcal{G}_\kappa^{++}/(1+\varepsilon) \\ \ell &\to \kappa \quad \text{or} \quad \ell \to -\kappa - 1 \\ \omega &\to \omega \\ E &\to (\varepsilon^2-1)/2\hbar^2 - \omega^2(\kappa - 1/2) \end{aligned} \qquad (16)$$

The choice $\ell \to \kappa$ or $\ell \to -\kappa - 1$ depends on whether $\kappa > 0$ or $\kappa < 0$, respectively. On the other hand, the map for $\mathcal{G}_\kappa^{--}$ is as follows:

$$\begin{aligned} g_\ell &\to 2\mathcal{G}_\kappa^{--}/(1-\varepsilon) \\ \ell &\to \kappa - 1 \quad \text{or} \quad \ell \to -\kappa \\ \omega &\to \omega \\ E &\to (\varepsilon^2-1)/2\hbar^2 - \omega^2(\kappa + 1/2) \end{aligned} \qquad (17)$$

Similarly, the choice $\ell \to \kappa - 1$ or $\ell \to -\kappa$ depends on whether $\kappa$ is positive or negative, respectively. Now, the nonrelativistic radial Green's function for the harmonic oscillator is well known [9]. It could be written as

$$g_\ell(r,r',E) = \frac{\Gamma(\frac{2\ell+3}{4} - E/2\omega^2)}{\omega^2 \Gamma(\ell+3/2)} \frac{1}{\sqrt{rr'}} \mathcal{M}_{E/2\omega^2,\frac{2\ell+1}{4}}(\omega^2 r_<^2)\mathcal{W}_{E/2\omega^2,\frac{2\ell+1}{4}}(\omega^2 r_>^2) \qquad (18)$$

where $\Gamma$ is the gamma function, $\mathcal{M}_{a,b}$ and $\mathcal{W}_{a,b}$ are the Whittaker functions of the first and second kind, respectively [10]. The two mappings (16) and (17) transform this nonrelativistic Green's function into the following solutions of Eq. (14):

$$\mathcal{G}_\kappa^{++} = \frac{1+\varepsilon}{2\omega^2}\frac{1}{\sqrt{rr'}}\begin{cases}\frac{\Gamma(-\mu+2\nu)}{\Gamma(2\nu+1)}\mathcal{M}_{\mu-\nu+\frac{1}{2},\nu}(\omega^2 r_<^2)\mathcal{W}_{\mu-\nu+\frac{1}{2},\nu}(\omega^2 r_>^2) &, \kappa>0 \\ \frac{\Gamma(-\mu)}{\Gamma(-2\nu+1)}\mathcal{M}_{\mu-\nu+\frac{1}{2},-\nu}(\omega^2 r_<^2)\mathcal{W}_{\mu-\nu+\frac{1}{2},-\nu}(\omega^2 r_>^2) &, \kappa<0\end{cases} \quad (19)$$

$$\mathcal{G}_\kappa^{--} = \frac{1-\varepsilon}{2\omega^2}\frac{1}{\sqrt{rr'}}\begin{cases}\frac{\Gamma(-\mu+2\nu)}{\Gamma(2\nu)}\mathcal{M}_{\mu-\nu,\nu-\frac{1}{2}}(\omega^2 r_<^2)\mathcal{W}_{\mu-\nu,\nu-\frac{1}{2}}(\omega^2 r_>^2) &, \kappa>0 \\ \frac{\Gamma(-\mu+1)}{\Gamma(-2\nu+2)}\mathcal{M}_{\mu-\nu,-\nu+\frac{1}{2}}(\omega^2 r_<^2)\mathcal{W}_{\mu-\nu,-\nu+\frac{1}{2}}(\omega^2 r_>^2) &, \kappa<0\end{cases} \quad (20)$$

where $\mu=(\varepsilon^2-1)/4\lambdabar^2\omega^2$ and $\nu=(\kappa+\frac{1}{2})/2$. The off-diagonal elements of $\mathcal{G}_\kappa$ are obtained by substituting these in Eq. (13), which could be rewritten as

$$\mathcal{G}_\kappa^{-+}(r,r',\varepsilon)=\mathcal{G}_\kappa^{+-}(r',r,\varepsilon)=\xi\frac{\lambdabar}{1+\varepsilon}\frac{1}{\sqrt{rr'}}\left(\frac{d}{dr}+\frac{\kappa-\frac{1}{2}}{r}+\omega^2 r\right)\sqrt{rr'}\mathcal{G}_\kappa^{++}$$
$$+(1-\xi)\frac{\lambdabar}{1-\varepsilon}\frac{1}{\sqrt{rr'}}\left(\frac{d}{dr'}-\frac{\kappa+\frac{1}{2}}{r'}-\omega^2 r'\right)\sqrt{rr'}\mathcal{G}_\kappa^{--} \quad (21)$$

Using the differential properties of the Whittaker functions [10] we obtain relations (A1) and (A2) in the Appendix, which when used in Eq. (21) give

$$\mathcal{G}_\kappa^{-+}(r,r',\varepsilon)=\mathcal{G}_\kappa^{+-}(r',r,\varepsilon)=(2\xi-1)(\lambdabar/\omega)\frac{\Gamma(-\mu+2\nu)}{\Gamma(2\nu)}\frac{1}{\sqrt{rr'}}\times$$
$$\left[\theta(r'-r)\mathcal{M}_{\mu-\nu,\nu-\frac{1}{2}}(\omega^2 r^2)\mathcal{W}_{\mu-\nu+\frac{1}{2},\nu}(\omega^2 r'^2)\right. \quad ,\kappa>0 \quad (22)$$
$$\left.+\frac{\mu}{2\nu}\theta(r-r')\mathcal{M}_{\mu-\nu+\frac{1}{2},\nu}(\omega^2 r'^2)\mathcal{W}_{\mu-\nu,\nu-\frac{1}{2}}(\omega^2 r^2)\right]$$

$$\mathcal{G}_\kappa^{-+}(r,r',\varepsilon)=\mathcal{G}_\kappa^{+-}(r',r,\varepsilon)=(2\xi-1)(\lambdabar/\omega)\frac{\Gamma(-\mu+1)}{\Gamma(-2\nu+2)}\frac{1}{\sqrt{rr'}}\times$$
$$\left[\theta(r'-r)\mathcal{M}_{\mu-\nu,-\nu+\frac{1}{2}}(\omega^2 r^2)\mathcal{W}_{\mu-\nu+\frac{1}{2},-\nu}(\omega^2 r'^2)\right. \quad ,\kappa<0 \quad (23)$$
$$\left.+(2\nu-1)\theta(r-r')\mathcal{M}_{\mu-\nu+\frac{1}{2},-\nu}(\omega^2 r'^2)\mathcal{W}_{\mu-\nu,-\nu+\frac{1}{2}}(\omega^2 r^2)\right]$$

where $\xi\neq\frac{1}{2}$. It is worthwhile noting that the resulting representation of the relativistic Green's function, as given above in Eqs. (20-23), has the advantage of computational economy as depicted in the number of terms – single terms. Another observation, although might be obvious, is that in the nonrelativistic limit ($\lambdabar\to 0$, $\varepsilon\to 1+\lambdabar^2 E$) the off-diagonal elements go to the limit like $\lambdabar$, whereas the lower diagonal element $\mathcal{G}_\kappa^{--}$ goes like $\lambdabar^2$.

**Dirac-Coulomb Green's function**: For review and recent examples on the use of the Dirac-Coulomb Green's function in the calculation of various physical processes one may consult Refs. [11] and references therein. In our approach for solving the relativistic wave equation, the Dirac-Coulomb problem is identified by taking $V(r)=Z/r$ and $W(r)=0$ [1,4,5]. This gives $S=\pm\lambdabar Z/\kappa$ where $\lambdabar Z=\alpha\mathbb{Z}$, $\alpha$ is the fine structure constant and $\mathbb{Z}$ is the dimensionless spinor charge in units of $e$. Consequently, Eq. (11) for the diagonal components of the radial Green's function reads as follows:

$$\left[-\frac{d^2}{dr^2}+\frac{\gamma(\gamma\pm 1)}{r^2}+2\frac{Z\varepsilon}{r}-\frac{\varepsilon^2-1}{\lambdabar^2}\right]\mathcal{G}_\kappa^{\pm\pm}(r,r',\varepsilon)=-(\gamma/\kappa\pm\varepsilon)\delta(r-r') \quad (24)$$

where $\gamma \equiv C\kappa = \sqrt{\kappa^2 - \lambdabar^2 Z^2} = \sqrt{\kappa^2 - \alpha^2 \mathbb{Z}^2}$ is the relativistic angular momentum. We compare this with the equation for the radial Green's function of the nonrelativistic Coulomb problem that reads

$$\left[ -\frac{d^2}{dr^2} + \frac{\ell(\ell+1)}{r^2} + 2\frac{Z}{r} - 2E \right] g_\ell(r,r',E) = -2\delta(r-r') \tag{25}$$

As a result of the comparison we obtain, by correspondence, the following two maps between the relativistic and nonrelativistic problems. The map for $\mathcal{G}_\kappa^{++}$ is

$$\begin{aligned} g_\ell &\to 2\mathcal{G}_\kappa^{++} / (\gamma/\kappa + \varepsilon) \\ \ell &\to \gamma \quad \text{or} \quad \ell \to -\gamma - 1 \\ Z &\to Z\varepsilon \\ E &\to (\varepsilon^2 - 1)/2\lambdabar^2 \end{aligned} \tag{26}$$

Again, the choice of either $\ell \to \gamma$ or $\ell \to -\gamma - 1$ depends on whether $\kappa > 0$ or $\kappa < 0$, respectively. The map for $\mathcal{G}_\kappa^{--}$, on the other hand, is as follows:

$$\begin{aligned} g_\ell &\to 2\mathcal{G}_\kappa^{--} / (\gamma/\kappa - \varepsilon) \\ \ell &\to \gamma - 1 \quad \text{or} \quad \ell \to -\gamma \\ Z &\to Z\varepsilon \\ E &\to (\varepsilon^2 - 1)/2\lambdabar^2 \end{aligned} \tag{27}$$

The choice $\ell \to \gamma - 1$ or $\ell \to -\gamma$ depends, as well, on whether $\kappa$ is positive or negative. Now, the nonrelativistic radial Green's function for the Coulomb problem is well known [12]. We write it as follows

$$g_\ell(r,r',E) = \frac{1}{2\sqrt{-2E}} \frac{\Gamma(\ell+1-\tau)}{\Gamma(2\ell+2)} \mathcal{M}_{\tau,\ell+\frac{1}{2}}(2\sqrt{-2E}\, r_<) \mathcal{W}_{\tau,\ell+\frac{1}{2}}(2\sqrt{-2E}\, r_>) \tag{28}$$

where $\tau = \sqrt{Z^2/-2E}$. The maps in (26) and (27) transform this into the following diagonal elements of the radial Dirac-Coulomb Green's function:

$$\mathcal{G}_\kappa^{++} = \frac{\gamma/\kappa + \varepsilon}{2\zeta} \begin{cases} \frac{\Gamma(\gamma-\mu+1)}{\Gamma(2\gamma+2)} \mathcal{M}_{\mu,\gamma+\frac{1}{2}}(\zeta r_<) \mathcal{W}_{\mu,\gamma+\frac{1}{2}}(\zeta r_>) &, \kappa > 0 \\ \frac{\Gamma(-\gamma-\mu)}{\Gamma(-2\gamma)} \mathcal{M}_{\mu,-\gamma-\frac{1}{2}}(\zeta r_<) \mathcal{W}_{\mu,-\gamma-\frac{1}{2}}(\zeta r_>) &, \kappa < 0 \end{cases} \tag{29}$$

$$\mathcal{G}_\kappa^{--} = \frac{\gamma/\kappa - \varepsilon}{2\zeta} \begin{cases} \frac{\Gamma(\gamma-\mu)}{\Gamma(2\gamma)} \mathcal{M}_{\mu,\gamma-\frac{1}{2}}(\zeta r_<) \mathcal{W}_{\mu,\gamma-\frac{1}{2}}(\zeta r_>) &, \kappa > 0 \\ \frac{\Gamma(-\gamma-\mu+1)}{\Gamma(-2\gamma+2)} \mathcal{M}_{\mu,-\gamma+\frac{1}{2}}(\zeta r_<) \mathcal{W}_{\mu,-\gamma+\frac{1}{2}}(\zeta r_>) &, \kappa < 0 \end{cases} \tag{30}$$

where $\mu = \lambdabar\sqrt{Z^2\varepsilon^2/(1-\varepsilon^2)}$ and $\zeta = (2/\lambdabar)\sqrt{1-\varepsilon^2}$. This representation has the same advantage of computational economy in the number of terms as that of the Dirac-Oscillator Green's function. The off-diagonal elements, on the other hand, are obtained by substituting these into Eq. (13) which, in this case, reads as follows:

$$\mathcal{G}_\kappa^{-+}(r,r',\varepsilon) = \mathcal{G}_\kappa^{+-}(r',r,\varepsilon) = \xi \frac{\lambdabar}{\gamma/\kappa + \varepsilon}\left(-\frac{Z}{\kappa} + \frac{\gamma}{r} + \frac{d}{dr}\right)\mathcal{G}_\kappa^{++}$$

$$+ (1-\xi) \frac{\lambdabar}{\gamma/\kappa - \varepsilon}\left(+\frac{Z}{\kappa} - \frac{\gamma}{r'} + \frac{d}{dr'}\right)\mathcal{G}_\kappa^{--} \tag{31}$$

Using the differential formulas and recurrence relations of the Whittaker functions [10] we obtain relations (A3) and (A4) of the Appendix which, when used in Eq. (31), give for $\kappa = 1,2,3\ldots$

$$\mathcal{G}_\kappa^{-+}(r,r',\varepsilon) = \mathcal{G}_\kappa^{+-}(r',r,\varepsilon) = -\frac{\lambdabar Z}{\gamma}\left[(\xi-1)\mathcal{G}_\kappa^{--} + \xi\frac{\gamma/\kappa-\varepsilon}{\gamma/\kappa+\varepsilon}\mathcal{G}_\kappa^{++}\right]$$
$$+\lambdabar(\xi-\tfrac{1}{2})\frac{\Gamma(\gamma-\mu+1)}{\Gamma(2\gamma+2)}\Big[\theta(r'-r)\mathcal{M}_{\mu,\gamma-\frac{1}{2}}(\zeta r)\mathcal{W}_{\mu,\gamma+\frac{1}{2}}(\zeta r') \qquad , \kappa>0 \qquad (32)$$
$$-\frac{\gamma+\mu}{2\gamma}\theta(r-r')\mathcal{M}_{\mu,\gamma+\frac{1}{2}}(\zeta r')\mathcal{W}_{\mu,\gamma-\frac{1}{2}}(\zeta r)\Big]$$

while, for negative values of $\kappa$, the result is

$$\mathcal{G}_\kappa^{-+}(r,r',\varepsilon) = \mathcal{G}_\kappa^{+-}(r',r,\varepsilon) = -\frac{\lambdabar Z}{\gamma}\left[(\xi-1)\mathcal{G}_\kappa^{--} + \xi\frac{\gamma/\kappa-\varepsilon}{\gamma/\kappa+\varepsilon}\mathcal{G}_\kappa^{++}\right]$$
$$+\lambdabar(\xi-\tfrac{1}{2})\frac{\Gamma(-\gamma-\mu+1)}{\Gamma(-2\gamma+2)}\Big[\frac{\gamma-\mu}{2\gamma}\theta(r'-r)\mathcal{M}_{\mu,-\gamma+\frac{1}{2}}(\zeta r)\mathcal{W}_{\mu,-\gamma-\frac{1}{2}}(\zeta r') \qquad , \kappa<0 \qquad (33)$$
$$+(2\gamma-1)\theta(r-r')\mathcal{M}_{\mu,-\gamma-\frac{1}{2}}(\zeta r')\mathcal{W}_{\mu,-\gamma+\frac{1}{2}}(\zeta r)\Big]$$

where $\xi \neq \tfrac{1}{2}$. Therefore, these off-diagonal elements do not share the advantage of economy in the number of terms with the diagonal ones. However, it should be noted that the expression inside the first brackets in Eq. (32) and Eq. (33) is second order in $\lambdabar$ resulting in third order relativistic terms while the remaining term is linear in $\lambdabar$. This could be shown by taking the nonrelativistic limit $\lambdabar \to 0$ which gives

$$\varepsilon \approx 1 + \lambdabar^2 E \quad , \quad \mathcal{G}_\kappa^{++} \approx g_\ell \quad , \quad \mathcal{G}_\kappa^{--} \approx \lambdabar^2 g_{\ell-1}$$
$$\gamma \approx \kappa - \lambdabar^2(Z^2/2\kappa) \quad , \quad \mu \approx \sqrt{Z^2/-2E} \quad , \quad \zeta \approx 2\sqrt{-2E} \qquad (34)$$

Using these limits, one can easily examine the relativistic behavior of the 2×2 radial Green's function which could be written symbolically as

$$\mathcal{G}_\kappa \sim \begin{pmatrix} 1 & \lambdabar+\lambdabar^3 \\ \lambdabar+\lambdabar^3 & \lambdabar^2 \end{pmatrix} \qquad (35)$$

It is very tempting to see whether this representation, which sieves out the third order relativistic component $\begin{pmatrix} 0 & \lambdabar^3 \\ \lambdabar^3 & 0 \end{pmatrix}$, induces any improvements in the calculation for a given physical process. It is, however, unfortunate that this author does not have the calculation tools nor the necessary computational skills needed to carry out such a project, which may prove to be very fruitful. One final remark to be made concerning the constant parameter $\xi$ that appears in the expressions for $\mathcal{G}_\kappa^{\pm\mp}$. Aside from the restriction that $\xi \neq \tfrac{1}{2}$, we found no obvious criterion for the selection of a specific value to be assigned to $\xi$. This issue might be settled based on the outcome of any possible improvements in the relativistic calculations due to this representation.

**Acknowledgments**: The author is indebted to Dr. Eric-Olivier Le Bigot for a stimulating exchange of messages, which motivated this work, and for the valuable comments on the original version of the manuscript. He is also grateful to Dr. M. I. Al-Suwaiyel and KACST Library for the valuable support in literature survey.

# Appendix

The following are some useful relations which could be obtained using the differential formulas and recurrence relations of the Whittaker functions [10]:

$$\left(\frac{d}{dx} + \frac{2b-1}{x} \pm x\right)\mathcal{M}_{a,b}(x^2) = 4b\,\mathcal{M}_{a\mp\frac{1}{2},b-\frac{1}{2}}(x^2)$$
$$\left(\frac{d}{dx} - \frac{2b+1}{x} \pm x\right)\mathcal{M}_{a,b}(x^2) = \left(\pm 1 - \frac{a}{b+\frac{1}{2}}\right)\mathcal{M}_{a\mp\frac{1}{2},b+\frac{1}{2}}(x^2)$$
(A1)

$$\left(\frac{d}{dx} - \frac{1\pm 2b}{x} + x\right)\mathcal{W}_{a,b}(x^2) = 2(a \mp b - \tfrac{1}{2})\mathcal{W}_{a-\frac{1}{2},b\pm\frac{1}{2}}(x^2)$$
$$\left(\frac{d}{dx} - \frac{1\pm 2b}{x} - x\right)\mathcal{W}_{a,b}(x^2) = -2\mathcal{W}_{a+\frac{1}{2},b\pm\frac{1}{2}}(x^2)$$
(A2)

$$\left(\frac{d}{dx} + \frac{b-\frac{1}{2}}{x} - \frac{a}{2b-1}\right)\mathcal{M}_{a,b}(x) = 2b\,\mathcal{M}_{a,b-1}(x)$$
$$\left(\frac{d}{dx} - \frac{b+\frac{1}{2}}{x} + \frac{a}{2b+1}\right)\mathcal{M}_{a,b}(x) = \frac{1/8}{b+1}\left[1 - \left(\frac{a}{b+\frac{1}{2}}\right)^2\right]\mathcal{M}_{a,b+1}(x)$$
(A3)

$$\left(\frac{d}{dx} + \frac{b-\frac{1}{2}}{x} - \frac{a}{2b-1}\right)\mathcal{W}_{a,b}(x) = -\frac{1}{2}\left(1 + \frac{a}{b-\frac{1}{2}}\right)\mathcal{W}_{a,b-1}(x)$$
$$\left(\frac{d}{dx} - \frac{b+\frac{1}{2}}{x} + \frac{a}{2b+1}\right)\mathcal{W}_{a,b}(x) = \frac{1}{2}\left(-1 + \frac{a}{b+\frac{1}{2}}\right)\mathcal{W}_{a,b+1}(x)$$
(A4)

**References:**

[1]  A. D. Alhaidari, Phys. Rev. Lett. **87**, 210405 (2001); **88**, 189901 (2002)

[2]  A. D. Alhaidari, J. Phys. A **34**, 9827 (2001); **35**, 6207 (2002)

[3]  A. D. Alhaidari, Int. J. Mod. Phys. A **17**, 4551 (2002)

[4]  A. D. Alhaidari, Phys. Rev. A **65**, 042109 (2002); **66**, 019902 (2002)

[5]  A. D. Alhaidari, "*Solution of the Dirac equation for potential interaction,*" Int. J. Mod. Phys. A, invited review article (in production)

[6]  G. Jian-You, F. Xiang Zheng, and Xu Fu-Xin, Phys. Rev. A **66**, 062105 (2002)

[7]  M. Moshinsky and A. Szczepaniak, J. Phys. A **22**, L817 (1989)

[8]  See, for example, J. Bentez, R. P. Martinez-y-Romero, H. N. Nunez-Yepez, and A. L. Salas-Brito, Phys. Rev. Lett. **64**, 1643 (1990); O. L. de Lange, J. Phys. A **24**, 667 (1991); V. M. Villalba, Phys. Rev. A **49**, 586 (1994); P. Rozmej and R. Arvieu, J. Phys. A **32**, 5367 (1999); R. Szmytkowski and M. Gruchowski, J. Phys. A **34**, 4991 (2001)

[9]  See, for examples, J. Bellandi Filho and E. S. Caetano Neto, Lett. Al Nuovo Cimento **16**, 331 (1976); E. Capelas De Oliveira, Revista Bras. de Fisica **9**, 697 (1979); J. H. Macek, S. Yu. Ovchinnikov and D. B. Khrebtukov, Rad. Phys. Chem. **59**, 149 (2000)

[10]  W. Magnus, F. Oberhettinger, and R. P. Soni, *Formulas and Theorems for the Special Functions of Mathematical Physics*, 3$^{rd}$ edition (Springer-Verlag, New York, 1966); H. Buchholz, *The Confluent Hypergeometric Function* (Springer-Verlag, New York, 1969); I. S. Gradshtein and I. M. Ryzhik, *Table of Integrals, Series and Products* (Academic, New York, 1980); H. Bateman and A. Erdélyi, *Higher Transcendental Functions* (McGraw-Hill, New York, 1953).

[11]  See, for examples, V. A. Yerokhin and V. M. Shabaev, Phys. Rev. A **60**, 800 (1999); P. J. Mohr, G. Plunien, and G. Soff, Phys. Rep. **293**, 227 (1998); C. Szymanowski, V. Véniard, R. Taïeb, and A. Maquet, Phys. Rev. A **56**, 700 (1997); D. J. Hylton and N. J. Snyderman, Phys. Rev. A **55**, 2651 (1997); R. A. Swainson and G. W. F. Drake, J. Phys. A **24**, 95 (1991); **24**, 1801 (1991).

[12]  See, for examples, L. Hostler and R. H. Pratt, Phys. Rev. Lett. **10**, 469 (1963); L. Hostler, J. Math. Phys. **5**, 591 (1964); **5**, 1235 (1964); **11**, 2966 (1970); A. Maquet, Phys. Rev. A **15**, 1088 (1977); B. R. Johnson and J. O. Hirschfelder, J. Math. Phys. **20**, 2484 (1979); S. M. Blinder, J. Math. Phys. **22**, 306 (1981); MKF. Wong and EHY. Yeh, J. Math. Phys. **26**, 1701 (1985)